\documentclass[authoryear, 12pt]{article}

\usepackage{amssymb}
\usepackage[document]{ragged2e}
\usepackage{graphicx}
\usepackage[table,xcdraw]{xcolor}
\usepackage{lscape}
\usepackage{adjustbox}
\usepackage{comment}
\usepackage{amsmath} 
\usepackage{tablefootnote}
\usepackage{graphics}
\usepackage{epsfig}

\usepackage{amssymb}

\usepackage{amsthm}
\usepackage[colorlinks = true,
           linkcolor = blue,
            urlcolor  = blue,
            citecolor = blue,
            anchorcolor = blue]{hyperref}
\usepackage{url} 
\usepackage{natbib}
\setcitestyle{aysep={,}} 

\usepackage{ragged2e}
\usepackage{makecell}
\usepackage{authblk} 

\usepackage[left=1.5cm, right=1.5cm, top=3cm, bottom=3cm, footskip=0.5cm]{geometry}

\definecolor{headercolor}{gray}{0.4}
\definecolor{lightgray}{gray}{0.9}

\author[1,2]{Grazia Sveva Ascione\thanks{Corresponding author: \texttt{graziasveva.ascione@uninsubria.it}. This research was funded by CNRS as part of the 2022-2023 "Programme Prématuration". We thank Andrea Vezzulli and participants of the RISE Workshop 2023 for their precious suggestions.}}
\author[1]{Valerio Sterzi}

\affil[1]{Bordeaux School of Economics, UMR CNRS 6060, University of Bordeaux}
\affil[2]{Department of Economics - DiECO, Varese, Università degli Studi dell'Insubria}
\date{}

\title{Presenting \textit{Terrorizer}: an algorithm for consolidating company names in patent assignees}
\begin{document}
\maketitle

\begin{abstract}
\justifying
 The problem of disambiguation of company names poses a significant challenge in extracting useful information from patents. This issue biases research outcomes as it mostly underestimates the number of patents attributed to companies, particularly multinational corporations which file patents under a plethora of names, including alternate spellings of the same entity and, eventually, companies' subsidiaries. To date, addressing these challenges has relied on labor-intensive dictionary based or string matching approaches, leaving the problem of patents' assignee harmonization on large datasets mostly unresolved. To bridge this gap, this paper describes the \textit{Terrorizer} algorithm, a text-based algorithm that leverages natural language processing (NLP), network theory, and rule-based techniques to harmonize the variants of company names recorded as patent assignees. In particular, the algorithm follows the tripartite structure of its antecedents, namely parsing, matching and filtering stage, adding an original "knowledge augmentation" phase which is used to enrich the information available on each assignee name. We use Terrorizer on a set of 325'917 companies' names who are assignees of patents granted by the USPTO from 2005 to 2022. The performance of Terrorizer is evaluated on four gold standard datasets. This validation step shows us two main things: the first is that the performance of Terrorizer is similar over different kind of datasets, proving that our algorithm generalizes well. Second, when comparing its performance with the one of  the algorithm currently used in PatentsView for the same task \citep{monathdisambiguating}, it achieves a higher F1 score. Finally, we use the Tree-structured Parzen Estimator (TPE) optimization algorithm for the hyperparameters' tuning. Our final result is a reduction in the initial set of names of over 42\%. 

\end{abstract}

\justifying

    \bigskip


\bigskip


\section{Introduction}

Patent data represent a significant source of information on innovation, knowledge production, and the evolution of technology through networks of citations, co-invention and co-assignment \citep{griliches1998patent, abraham2001innovation, hall2012recent, sampat2018survey}.  Patent related indicators are now used by companies and by policymakers and governmental agencies alike to assess technological progress and knowledge diffusion on the level of regions, countries, domains , and even specific entities such as companies, universities and individual inventors \citep{pavitt1985patent, von2005inventive, van2006data, squicciarini2013measuring}. However, with respect to the patentee level, specific concerns can be discerned. Despite recent advancements in technology intelligence and patents' information retrieval, a major obstacle to extracting useful information from this data is the problem of name disambiguation: linking alternate spellings of individuals or organizations to a single identifier to uniquely determine the parties involved in knowledge production and diffusion \citep{magerman2006data, raffo2009play, carayol2009s, thoma2010harmonizing, peeters2010harmonizing, pezzoni2014kill, morrison2017disambiguation}. This problem biases the results of related research,
creating problems in identifying, for instance, prolific inventors \citep{gay2008collective} or in assessing the number of patents assigned to a single entity which files under a plethora of different names, such as multinational corporations \citep{magerman2006data}.
Therefore, the concern which stems from the heterogeneity  of patentee documents affects not only researchers, but also complicate the innovation analysis for companies and government  stakeholders as well, who have troubles correctly identifying patent inventors and assignees.
Although the problem of inventors' disambiguation has been extensively analyzed in literature (i.e. \cite{trajtenberg2006names,  carayol2009s, lai2009careers, raffo2009play, miguelez2011singling, li2014disambiguation,  pezzoni2014kill, ventura2015seeing, kim2016inventor, morrison2017disambiguation, han2019disambiguating, petrie2023novel}), only few works deal with disambiguation of company names \citep{magerman2006data, peeters2010harmonizing, thoma2010harmonizing, neuhausler2017identifying, lee2023simple}.
In particular, disambiguating company names has been linked to a specific set of problems, including spelling and spaces mistakes, dealing with different languages, deleting keywords related to the legal form of a certain company, as well as geographical indications and company which change names over time \citep{magerman2006data, peeters2010harmonizing}. Therefore, in the same dataset it is possible to find a wide range of alternate spellings of a single company name and also its subsidiaries.\footnote{For instance, the company "NOKIA" appears in 89 variations in USPTO patent applications, including spelling mistakes and subsidiaries such as "NOKIA MOBILES PHONES, LTD.", "NOKIA TECHNOLOGIES OY", "NOKIA US HOLDINGS INC." and so on. In addition, the order of tokens does not always see the first token as the company proper names, such as in the case of "BEIJING SAMSUNG TELECOM R\&D CENTER" referring to "SAMSUNG".}
To date, these problems have often been tackled with labor-intensive and hard-coded approaches such as language specific dictionaries or with string matching procedures, leaving the question of how to solve the patents' assignees harmonization task on large datasets unsolved. 
To fill this gap, in this paper we present the \textit{Terrorizer} algorithm, an original text-based algorithm aimed at harmonizing the variants of the companies’ names which are registered as patents’ assignees in patents' databases relying on a mixture of natural language processing (NLP), network theory and rule-based techniques. An algorithm as such could provide significant benefit to the community of scholars working on patent data for the resulting high quality assignee disambiguation.
We use Terrorizer to perform entity linking on USPTO assignees and applicants of granted utility patents from 2005 to 2022. We retrieve the data from both the Patent Assignment Database (PAD, Version 2021) that includes all the assignee names in the patent historyand from PatentsView to retrieve original patent assignees. Our focus is on company names and our algorithm shows encouraging results in this sense. Considering the whole sample of 325'917 company names, we  reduce unique names to 188'445 (around 42\%).
Furthermore, we assess the performance of Terrorizer exploiting four different datasets with groundtruth: two come from the research of \cite{monathdisambiguating} and allow us to compare the performance of Terrorizer to their algorithm for cleaning patent assignees' names; other two are created \textit{ad hoc} for this research using PAD data. Our validation results show that the performance of Terrorizer is stable across the four datasets, pointing out that this algorithm is good at generalizing. Further, the performance in terms of F1 score is better on both groundtruth datasets compared to the F1 achieved by the algorithm of \cite{monathdisambiguating}.
Beyond that, it is important to underline that the performance of the algorithm varies according to the used parameters and that one could decide to favor precision over recall or vice-versa instead of their harmonic mean (F1 score). For this reason, we perform hyperparameter optimization using the Tree-structured Parzen Estimator (TPE), with the aim of maximizing the F1 score. Eventually, we use the optimized hyperparameters of one of the benchmark datasets to do a new run of Terrorizer and achieve the final results. 
Finally, this paper adds to the literature proposing a novel way of exploiting recent NLP advances and network theory to solve an economically relevant problem. The use of such a sophisticated name matching strategies facilitates a more accurate view of patent portfolios of agents or institutions. It can therefore significantly modify the results on inventive activity, technological profiles or networking of the companies involved in patent production \citep{raffo2009play}.
The rest of the paper is organized as follows. Section \ref{section2} discusses the state-of-the-art on the matter, Section \ref{sec:meth} illustrates the proposed methodology. Section \ref{sec:res} presents the results, and Section \ref{sec:co} concludes.
\clearpage
\section{State-of-the-art}\label{section2}

Several names disambiguation's approaches have been developed to harmonize the different name variants occurring for the same company, spanning from dictionary based methods to deep-learning based approaches. In this section we describe the most relevant ones.\footnote{Beyond methodological papers, there have been other efforts to reduce the variety of patent assignees names. For instance, in 2019, the UVA DARDEN School of Business (University of Virginia) funded the construction of a new patent-firm linked database: Global Corporate Patent
Dataset. The dataset covers patents awarded by the USPTO to
publicly listed firms internationally
between 1980 and 2017. Their harmonization methodology proposes a combination of string matching and knowledge augmentation techniques using companies websites' urls and stock information available in search engines. The full description of the dataset and of the linked harmonization methodology is available at \url{https://patents.darden.virginia.edu/documents/DataConstructionDetails_v01.pdf}. Moreover, we omit in this Section to comment on the papers which perform name harmonization but not as primary goal of the paper itself (see for instance, recent work by \cite{arora2021matching}).} Before presenting the different methodologies, it is important to address the diverse problems that the literature highlights regarding the harmonization of assignees' names in patents, especially when they pertain to companies:

\begin{enumerate}
    \item Spelling variations (different but correct spelling variations, such as "Bain \& co." and "Bain and company") \citep{magerman2006data, thoma2010harmonizing}; \label{point1}
    \item Spelling mistakes (such as "Nokia international" and "Nokia interational") \citep{magerman2006data, thoma2010harmonizing, onishi2012standardization, li2014disambiguation}; \label{point2}
    \item Business and legal extensions (such as "IBM LTC." and "IBM INC.") \citep{magerman2006data, lee2023simple}; \label{point3}
    \item Addition of geographical indication   (such as "BASF Europe" and "BASF Beijing")  \citep{magerman2006data}; \label{point4}
    \item Identification of subsidiaries and of ownership changes (such as "Nokia Bell Labs" and "Nokia Siemens Network" which both are subsidiaries of Nokia), defined by \cite{magerman2006data} as \textit{"legal entity harmonization"} \citep{li2014disambiguation};\label{point5}
    \item Acronyms (such as "IBM" for "International Business Machines") \citep{magerman2006data, li2014disambiguation, peeters2010harmonizing}; \label{point6}
\end{enumerate}

Table \ref{tab:literature} summarizes the different strategies adopted for the above mentioned task in chronological order, the type of harmonization proposed (\textit{internal} when it performs entity resolution within a single list; \textit{external} when it matches two different names' lists), the problems addressed, the data used for the experiments, the technical characteristics and the limitations. 
The methodology of \cite{magerman2006data} focuses on the identification of name variations by comparing each patentee name with all other patentee names; the main objective is to match names that appear to be similar but differ because of spelling or language variations, but without dealing with legal entity harmoniziation. Their approach consist in a pre-processing stage where they clean punctuation, convert the text to standard ASCII characters, and a name cleaning stage where they use dictionaries to clean legal extensions in the most important languages, manually change the most frequently occurring spelling mistakes and performing Umlaut harmonization for German characters with a diacritic mark. They apply their methodology on an integrated set of EPO and USPTO patentee names, reducing the number of unique patentee names by 17.6\%, from 443'722 to 365'866 names. However, this methodology is labour intensive for the necessity of manual checks, and requires constant updates when there are new sets of names to be harmonized; for these reasons, the harmonization pipeline is difficult to automatize. Moreover, the degree of comprehensiveness of such dictionaries strongly impacts both the precision and the recall of this methodology, considering, for the former, that some deleted words (such as geographical indications) might be proper names for some companies\footnote{For instance, it is appropriate to remove “France” in “ABB France” but not in “France
Telecom” \citep{peeters2010harmonizing}.}; while, for the latter, that common keywords not present in the dictionaries will lower the recall of the method. 

This methodology is commented on and improved by  \cite{peeters2010harmonizing} who, with further manual labor, attempt to enhance names' harmonization. Their goal is to maximize the impact of their methodology in terms of information which can be retrieved from patents; this is achieved by selecting the top 500 actors based on cumulative counts for
EPO/USPTO/WIPO patent documents. After, for names in the top 500 they perform a string similarity search (using the Levenshtein distance) to identify the possible variants. Despite the improvements of this approach compared to the one presented in \cite{magerman2006data}, this method  suffers of a number of limitations: first, the labor intensive manual correction step is fundamental, second string matching procedures do not lead to the desired results in many cases\footnote{For instance, the Levenshtein distance for cases such as "NOKIA SOLUTIONS AND NETWORKS MANAGEMENT INTERNATIONAL" and "NOKIA TELECOMMUNICATIONS OY" will give a high distance, while cases such as "IMTECH, INC." and "AMTECH, INC." will have a very low distance even if they are different companies.} and they are computationally inefficient on large datasets\footnote{The Levenshtein algorithm has a time complexity of \(O(m \times p)\), where \(m\) and \(p\) are the lengths of the two strings being compared. Therefore, the combined computational cost would be prohibitively high for large datasets, making it an impractical choice for this specific research context.}, therefore difficult to apply on sets of patents' assignees names.

\cite{thoma2010harmonizing} combine dictionary and rule-based approaches to consolidate European, Patent Cooperation Treaty (PCT) and US patent data with firm accounting data. The resulting data encompass about 131’000 patent applicant names, covering more than 50\% of EPO and PCT applications and also of USPTO assignees names. In particular, they expand the previous dictionary and string matching based methodologies using priority links across USPTO and EPO patent databases as an additional harmonization method. However, the extra information they combine with dictionary and string matching methodologies is limited to those assignees linked to one or more patent families.

\cite{onishi2012standardization} work on harmonization of company applicants at the Japanese Patent Office (JPO). They focus on 1’638 Japanese listed firms in the manufacturing, ICT and distribution industry using not only the applicants’ names, addresses, and a large directory database of Japanese firms, but also historical company profiles to identify name changes over the years. Their methodology consists of a manual collection of company data, including addresses as listed in financial reports and websites; in addition, in cases where the applicant names could not be accurately identified using the above two steps, direct inquiries were sent by e-mail to the most likely companies via email. They find 20’616 name variations in total for their companies of interest. Even if their approach is very accurate, it requires manual effort and therefore it is not scalable for bigger names datasets.

\cite{neuhausler2017identifying} match data on German R\&D expenditures with EPO patent data at the micro-level, i.e. at the level of companies and patent applicants, respectively. They build on data created by previous work by \cite{magerman2006data} and \cite{peeters2010harmonizing} (the EEE-pat dataset) and match it to company names present in the SV Wissenschaftsstatistik (German R\&D survey data). To do that, they first perform another round of text preprocessing (on top of the one previously performed) and then use a measure of normalized  Levenshtein distance (by the number of characters in the longer text-string) to do the match with the R\&D list. Further, to decide whether two names match, they also consider whether the first 3 digits of the  ZIP code are the same. If those are completely different, the potential match is discarded despite the textual similarity. Afterwards, they set an arbitrary threshold and they selected it using a manually cleaned sample of 1'000 randomly selected names from the R\&D dataset in order to maximize the F1 score. Limitations of this research include the fact that it is specific to that subset of German observations and that it does not allow matching of semantic distant subsidiaries.

\cite{monathdisambiguating} propose a disambiguation methodology which concerns inventors, locations and assignee names. Their overall strategy aims to identify which separate occurrences of an inventor, assignee, or location name (referred to as a mention) are the same person, organization, or location. The disambiguation process seeks to resolve two overlapping data issues:
i) multiple names for the same entity (inventor, assignee, or location); and
ii) multiple different entities with the same name. The first step is to group records (or, more precisely, mentions from the
patent records) into “canopies”. Canopies are formed according to similarity rules. For what concerns assignees names, they use an exact four-character overlap of the beginning of any word or name of the organization as the criteria for creating assignee canopies. After, they use a system of pairwise similarity scoring, which checks a series of conditions such having the same \textit{PERMid}\footnote{\url{https://permid.org/}} and other string matching attributes. Finally, they group together the two most similar records and each time records are
clustered together, they form a node in a tree. This step is repeated recursively. After all records are formed into a tree, the final clusters are a subtree whose similarity score exceeds a determined threshold. This text-based methodology would not work well when the names considered are semantically very distant, therefore does not tackle legal entity harmonization.

\cite{lee2023simple} propose instead a deep-learning based approach to disambiguate patent applicants. In particular, they propose an augmented data approach which trains an attention-based LSTDM model. In particular, in subsequent steps they respectively add noise to the original names using basic business entity specific lexicon, then they add punctuation and finally perform permutations among patent applicants names to increase the diversity of the dataset. After training the model, they perform two experiments on a dataset of 439 patent applicants and on another containing companies in “Fortune 500" 2022 list and companies listed on the Korean Composite Stock price Index (KOSPI). They obtain an accuracy of 0.96 and 0.94 respectively. However, the proposed method is heavily reliant on the dictionary used for the training set and it does not consider legal entity harmonization.

\begin{landscape}
\begin{table}[htbp]
\begin{center}
\caption{Proposed methodologies for companies' name harmonization}
\label{tab:literature}
\medskip
\setlength{\tabcolsep}{8pt} 
\renewcommand{\arraystretch}{1.1} 
\scalebox{0.6}{
\begin{tabular}{|l|l|l|l|l|l|l|}
\hline
{\cellcolor[HTML]{EFEFEF}{\color[HTML]{1F1F1F} \textbf{Author(s)}}} &
{\cellcolor[HTML]{EFEFEF}  {\color[HTML]{1F1F1F} \textbf{Year}}} &
{\cellcolor[HTML]{EFEFEF}  {\color[HTML]{1F1F1F} \textbf{Target data}}} &
{\cellcolor[HTML]{EFEFEF}  {\color[HTML]{1F1F1F} \textbf{Kind of match}}} &
{\cellcolor[HTML]{EFEFEF}  {\color[HTML]{1F1F1F} \textbf{\begin{tabular}[c]{@{}l@{}}Points \\ addressed\end{tabular}}}} &
{\cellcolor[HTML]{EFEFEF}  {\color[HTML]{1F1F1F} \textbf{Main technique(s)}}} &
{\cellcolor[HTML]{EFEFEF}  {\color[HTML]{1F1F1F} \textbf{Limitations}}} \\ \hline
{\color[HTML]{1F1F1F} \textbf{Magerman et al.}} &
  {\color[HTML]{1F1F1F} 2006} &
  {\color[HTML]{1F1F1F} EPO applicants and USPTO assignees} &
  {\color[HTML]{1F1F1F} internal} &
  {\color[HTML]{1F1F1F} 1,2,3,4,6} &
  {\color[HTML]{1F1F1F} \begin{tabular}[c]{@{}l@{}}-text preprocessing\\ -creation of dictionaries\\ -rule-based techniques\end{tabular}} &
  {\color[HTML]{1F1F1F} \begin{tabular}[c]{@{}l@{}}-labor intensive\\ -does not deal with legal entity harmonization \\ (therefore low recall)\\ -precision and recall impacted by dictionaries’ \\ completeness\end{tabular}} \\ \hline
{\color[HTML]{1F1F1F} \textbf{Peeters et al.}} &
  {\color[HTML]{1F1F1F} 2010} &
  {\color[HTML]{1F1F1F} \begin{tabular}[c]{@{}l@{}}Top 500 patentees in EPO/USPTO/WIPO \\ from PATSTAT\end{tabular}} &
  {\color[HTML]{1F1F1F} internal} &
  {\color[HTML]{1F1F1F} 1,2,3,4,5,6} &
  {\color[HTML]{1F1F1F} \begin{tabular}[c]{@{}l@{}}-manual checks\\ - Levenshtein distance\end{tabular}} &
  {\color[HTML]{1F1F1F} \begin{tabular}[c]{@{}l@{}}-labor intensive\\ -missing all those cases where string matching \\ procedures return a high score\\ -high computational costs for those string \\ matching procedures\\ -legal entity harmonization performed\\  through  manual online searches\end{tabular}} \\ \hline
{\color[HTML]{1F1F1F} \textbf{Thoma et al.}} &
  {\color[HTML]{1F1F1F} 2010} &
  {\color[HTML]{1F1F1F} 131’000 patent applicant names} &
  {\color[HTML]{1F1F1F} \begin{tabular}[c]{@{}l@{}}internal and \\ external\end{tabular}} &
  {\color[HTML]{1F1F1F} /} &
  {\color[HTML]{1F1F1F} \begin{tabular}[c]{@{}l@{}}-matching of UPSTO and EPO assignees \\ through patent families\end{tabular}} &
  \cellcolor[HTML]{FFFFFF}{\color[HTML]{1F1F1F} - extra information is limited to patent families} \\ \hline
{\color[HTML]{1F1F1F} \textbf{Onishi et al.}} &
  {\color[HTML]{1F1F1F} 2013} &
  {\color[HTML]{1F1F1F} \begin{tabular}[c]{@{}l@{}}1’638 Japanese listed firms in the\\  manufacturing, ICT and distribution\\  industry\end{tabular}} &
  {\color[HTML]{1F1F1F} internal} &
  {\color[HTML]{1F1F1F} 1,2,3,4,5,6} &
  {\color[HTML]{1F1F1F} \begin{tabular}[c]{@{}l@{}}-hand collection of companies data including \\ addresses and/or the other below mentioned resources\\ -direct email to companies to solve ambiguities\end{tabular}} &
  {\color[HTML]{1F1F1F} \begin{tabular}[c]{@{}l@{}}-only based on manual labor\\ -they work on “single” companies \\ (not on legal entity harmonization)\end{tabular}} \\ \hline
{\color[HTML]{1F1F1F} \textbf{Neuhausler et al.}} &
  {\color[HTML]{1F1F1F} 2017} &
  {\color[HTML]{1F1F1F} German patent applicants from 2007 to 2009} &
  {\color[HTML]{1F1F1F} external} &
  {\color[HTML]{1F1F1F} 3} &
  {\color[HTML]{1F1F1F} \begin{tabular}[c]{@{}l@{}}-text preprocessing\\ - modified version of Levenshtein distance\\ -add a ZIP code criterion (where present)\end{tabular}} &
  {\color[HTML]{1F1F1F} \begin{tabular}[c]{@{}l@{}}-highly specific\\ -does not consider legal entity harmonization\end{tabular}} \\ \hline
  {\color[HTML]{1F1F1F} \textbf{Monath et al.}} &
  {\color[HTML]{1F1F1F} 2021} &
  {\color[HTML]{1F1F1F} all patent assignees from USPTO} &
  {\color[HTML]{1F1F1F} internal} &
  {\color[HTML]{1F1F1F} 1,2,3,6} &
  {\color[HTML]{1F1F1F} \begin{tabular}[c]{@{}l@{}}-creation of canopies\\ - similarity calculation within canopies\\ -hierarchical clustering\end{tabular}} &
  {\color[HTML]{1F1F1F} \begin{tabular}[c]{@{}l@{}}-does not consider when names are very different\\ -not suitable for legal entity harmonization\end{tabular}} \\ \hline
{\color[HTML]{1F1F1F} \textbf{Lee et al.}} &
  {\color[HTML]{1F1F1F} 2023} &
  {\color[HTML]{1F1F1F} \begin{tabular}[c]{@{}l@{}}439 patent applicants (D1) + list of Global\\  companies  and companies listed on KOSPI\end{tabular}} &
  {\color[HTML]{1F1F1F} internal} &
  {\color[HTML]{1F1F1F} 3} &
  {\color[HTML]{1F1F1F} \begin{tabular}[c]{@{}l@{}}-train an attention-based LSTM model\\  using an -augmented training dataset\\  created adding noise to company names.\end{tabular}} &
  {\color[HTML]{1F1F1F} \begin{tabular}[c]{@{}l@{}}-highly dependent on the training dataset\\ -semantic might  not enough to disambiguate \\ company names\\ -does not consider legal entity harmonization\end{tabular}} \\ \hline
\end{tabular}%
}
\end{center}
\end{table}
\scriptsize
{Notes: The table present a summarization of the different methodologies for the task of entity resolution on company names related to patent assignees. In particular, the first column reports the author(s), the second the year of publication, the third the kind of match (\textit{intra list} or between different lists of names). Further, the third column reports which of the points, described in Section \ref{section2}, are directly addressed in each of the papers, while the fourth column addresses the most important limitations.}

\end{landscape}

\section{Methodology: a "new" three (and a half) stage game}
\label{sec:meth}

\subsection*{Overview}

\begin{figure}
  \caption{The pipeline of Terrorizer}
\bigskip{}
  
  \centerline{\includegraphics[width=0.9\textwidth]{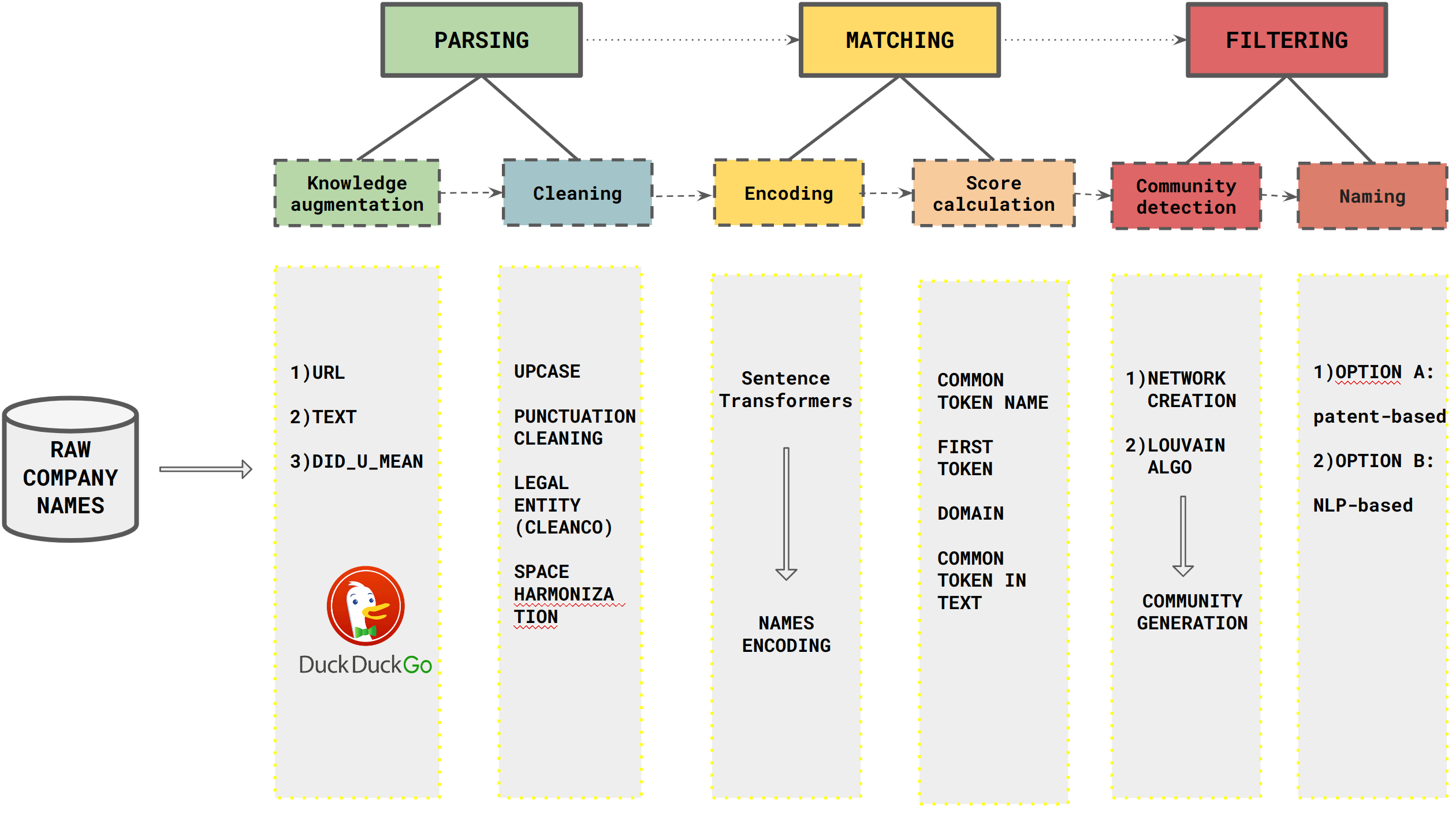}}
\scriptsize
  \justifying{\emph{Notes:} The figure reports the three main phases of Terrorizer and the six linked subphases: parsing (containing knowledge augmentation and cleaning), matching (including encoding and score calculation) and filtering (entailing community detection and naming).}
\label{fig:matching0}
\end{figure}
\bigskip{}

The goal of this work is to improve the previous performances in terms of companies' names harmonization in patent assignees. In Terrorizer, we combine network theory with a NLP and rule-based approach as the previously adopted approaches would not be able to be exploited for a number of reasons. The most important one is that dictionary based methods are not feasible on great volumes of data because of their intrinsic variety of the names. Furthermore, string similarity approaches are also not a feasible option on company names as well. First, because, as demonstrated with the above examples, the same company can be linked to names variations which have a low string similarity; for instance "NOKIA LONDON" will going to be more similar to "SAMSUNG LONDON" than to "NOKIA RESEARCH AND DEVELOPMENT OPERATIONS UK". Second, string similarity algorithms are computationally expensive applied on huge datasets. 
Terrorizer's approach to overcoming these problems is presented in the following subsections. We make use of the recent improvements brought by NLP and deep learning in particular, to the field of patent analytics \citep{li2018deeppatent, sarica2020technet, trappey2019machine, arts2021natural}. Figure \ref{fig:matching0} reports the pipeline of the proposed algorithm, which is composed by three stages  accordingly with the three stage approach already adopted by \cite{raffo2009play} and \cite{pezzoni2014kill}. First of all, we  start with the \textit{parsing} stage. In this step, we rely on a \textit{knowledge augmentation} phase to add information to the company names; this step involves the creation of a crawler which extracts information through a web search on each of the names and stores it in a database for later use; then, we proceed with a general cleaning of the names. We then compare each pair of names in our data and assign them a score in the \textit{matching} stage. In this stage, we also consider the cosine similarity between each pair of names as a part of the matching score. Finally, in the \textit{filtering} stage we form communities using the principles of network theory. Further detail on the functioning of the algorithm is given in the following subsections. In particular, subsection \ref{subsec1} presents the data used to test the algorithm; the three followings present the proposed methodology in its three stages: parsing and knowledge augmentation (subsection \ref{subsec2}), matching (subsection \ref{subsec3}) and filtering (subsection \ref{subsec4}). Finally, we present the validation procedure (subsection \ref{subsec:val}) and conclude the section with the hyperparameter optimization (subsection \ref{subsec:opt}).

\subsection{Data} \label{subsec1}
\bigskip{}

To test the performance of the Terrorizer algorithm, in this research we harmonize patent applicants and assignees names of utility patents granted at the USPTO. The aim is to create a database of all assignees of USPTO patents filed since 2005, including original applicants and subsequent buyers. Therefore, we combine two data sources: the first is the Patent Assignment Database (PAD, Version 2021), where we have the information on assignees buying patents, and the second is the \textit{Applicant} table from PatentsView (PW), where we have the original applicants. We retrieve from PAD unique assignees names linked to three conveyance types: "assignment", "government" and "merger".\footnote{In particular, we use the assignment\_conveyance table to identify the transaction kinds of interest, the assignment table as reference for the dates of the transactions and the assignor and assignee table to get the related assignor and assignees names. Amongst the conveyance
types, assignments and mergers are the most self-evident of a real change in ownership
where property transfers from one party to another or from one corporate entity to
another. Further, considering that the definition of \textit{government} assignment in \cite{graham2018patent} is "A government interest agreement is a license, assignment or other interest of the Federal
Government in or under a patent or patent application.", we consider this kind of assignment a real transaction and therefore we keep it in our analysis.
All transactions' related tables are freely available at \url{https://www.uspto.gov/ip-policy/economic-research/research-datasets/patent-assignment-dataset}}

Further, we consider only those names related to transactions which took place starting from 2005 onwards. Considering not all patents are transacted during their life, we add to our assignee list the names of applicants of patents, filed since 2005, which have never been transferred, using PW.\footnote{The name of the table is g\_applicant\_not\_disambiguated.tsv and it has retrieved at the following url \url{https://patentsview.org/download/data-download-tables}}
In total, we collect 325'917 different patent assignees' names linked to 3'354'209 patents.\footnote{To decide which names refer to companies and which to individuals we finetune DistilRoBERTa base \citep{sanh2019distilbert} using manually labeled data about patent assignee kind and add a final classification layer. The technical details of the classificator are presented in the Appendix \ref{sec:SectionA.3}; further, we identify universities and research institutions using a multilingual keyword list.} 

\subsection{Parsing phase: knowledge augmentation and cleaning} \label{subsec2}
\bigskip{}

The first step is to collect extra information on the companies' names. As mentioned in point \ref{point2} of Section \ref{section2}, correcting spelling mistakes is defined as a crucial issues for this kind of names cleaning. To do that, as well as to add valuable information to the companies’ names, we rely on \textit{data augmentation}, creating a web crawler which extracts data from the web search engine \textit{DuckDuckGo}.\footnote{DuckDuckGo is an internet search engine that emphasizes protecting searchers' privacy and avoiding the filter bubble of personalized search results} 

Using web search engines as a way to standardize patentee company names has been proposed by \cite{autor2020foreign}, who use the search engine Bing to improve the matching between Compustat and USPTO patent data. In this research, we expand this idea using web searches as a way not only to standardize but also to increase information available for each company name. In particular, we retrieve for each company name reported the correct spelling if present, the url and the corresponding text from the first result of the search output; this information is going to be used later in the \textit{matching} stage. In particular, the crawler automatically digits the previously parsed name as a query in the browser; to get the correct name, we rely on the information in the "\textit{did\_u\_mean}" html tag in the search engine results web page for each name, which reports, if the text entered as a query contains some errors, the correct spelling of that name. With this technique, we retrieve 20'461 names for which the \textit{did\_u\_mean} feature is present. Some examples of the corrected spelling mistakes include: "PHILIPS HEALTCARE INFORMATION INC." which is reported in the \textit{did\_u\_mean} tag as "PHILIPS HEALTHCARE INFORMATION", "INNOVASION LABS, INC." reported as "INNOVATION LABS, INC.", "CELLESTIS" corrected as "CELLECTIS" and so on.
After the augmented information is collected, we then start the parsing of the names. More in detail, the following pre-processing steps are performed: replacing the original name with the name stored in the \textit{did\_u\_mean} variable where present\footnote{It is important to note that we cannot be certain that, in all the cases, the name stored in the \textit{did\_u\_mean} tag is correct. However, comparing the results of DuckDuckGo with other search engines, such as Google or Bing, we find the former correct in the majority of cases. }, lower-casing, spaces stripping and punctuation signs removal. Then, the legal form in the most frequently occurring languages are deleted using the lists in the \textit{cleanco} python package which processes company names, providing cleaned versions of the names by stripping away terms indicating organization type (such as "Ltd." or "Corp").\footnote{\url{https://pypi.org/project/cleanco/}}

The pre-processing operations reduce the number of unique names to 259'856 (corresponding to a reduction of around 20\%). Further, the augmented information (url and text of the first url) previosuly collected is also parsed. In particular, the domain of the website is extracted from the url and only retained if it is not one of the 100 most frequent domains in the data. The reason for this is that we later want to compare pairs of names and increase the probability that they refer to the same entity if they have the same domain. Therefore, we want drop from the analysis domains which do not refer to specific companies but to more general sources of knowledge (including, for instance, news website, Wikipedia, online databases and so on). The text of the first url is also pre-processed.\footnote{After being cleaned from extra whitespaces and punctuation, we remove from each text all the words contained in our list of most frequently occurring words. This list has been realized picking the top 250 most frequently occurring words.} As anticipated, this information is later exploited in the \textit{matching} stage.

\subsection{Matching phase} \label{subsec3}
\bigskip{}

In the matching phase, for each couple of names in our data, we want to calculate a \textit{matching score}. To get the matching score for each pair, we verify if a series of four conditions are met and we also calculate the cosine similarity between the two names.\footnote{To calculate the cosine similarity we use the cosine similarity function inside the Sentence-Transformers package. Details on the implementation are available at \url{https://www.sbert.net/docs/package_reference/SentenceTransformer.html}.} In particular, for each couple of names, we calculate the scalar product between a vector where each condition and the cosine similarity represent a dimension of that vector and a vector of weights. A summary of the components of the former is provided in Table \ref{tab:vec}.

Before proceeding with the matching stage, exploiting the list of most common words created in the parsing stage, we verify whether a company name is composed of words that are all contained in the list of most occurring words or not. We create such a division because company names which contain only "generic" tokens are less likely to produce reliable results during the knowledge augmentation phase.\footnote{For instance, it is difficult to say whether the entity "PHARMA GROUP, INC." should be linked to Pharma Resource Group (first result from DuckDuckGO), Pharma Group in Miami (second result) or Ferndale Pharma Group (third result).} 
Consistently, we compute the matching score using vectors of different dimensions for the two categories of names, treating them separately. For \textit{type 1} names (314'856 names before preprocessing) we create a 5th dimensional vector including four zero/one elements, where we have 1 if the condition is met and 0 otherwise. The four conditions are as follows: 

\begin{itemize}
    \item the two names have a token in common (where the tokenization is done by spaces);
    \item the first token is in common, where the previous condition is met;
    \item the domain of the two names is the same;
    \item there is a common word in the texts of the url, after verifying for both names that there is at least one common word between each parsed name and its text;
\end{itemize}
\begin{table}[h]
\caption{Matching score conditions (\textit{type 1} names)}
\medskip
\label{tab:vec}
\renewcommand{\arraystretch}{1.5} 
\resizebox{\textwidth}{!}{  
\centering
\begin{tabular}{|l|l|c|c|}
\hline
\rowcolor{headercolor} \color{white}\textbf{Condition} & \color{white}\textbf{Description} & \color{white}\textbf{Possible Values} & \color{white}\textbf{Feature Weight \tablefootnote{\label{n1}In the first run of Terrorizer, all the weights are set to 1. After the hyperparameter optimization phase explained in Section \ref{sec:opt}, the assigned weights are those selected by the optimization process.}} \\
\hline
\rowcolor{lightgray} \textbf{Any token in common}&The two names have a token in common\tablefootnote{The tokenization is done by spaces.}  & 0 or 1 & 1 \\
\hline
\textbf{First token in common }&The two names have the first token in common& 0 or 1& 1 \\
\hline
\rowcolor{lightgray} \textbf{Common token in url's text} & There is a common token in the text of the url & 0 or 1 & 1 \\
\hline
\textbf{Domain in common} & The two names have the same domain & 0 or 1 & 1 \\
\hline
\rowcolor{lightgray} \textbf{Cosine similarity}& The cosine similarity between the two names & from -1 to 1 & 1 \\
\hline
\end{tabular}
}
\end{table}
{\scriptsize
Notes: The table present a summarization of the five conditions whose possible values compose the vector of conditions for \textit{type 1} names. In particular, the first column reports the condition itself, the second its description, the third possible values and the fourth reports the values composing the vector of weights.} 

\bigskip{}

Furthermore, we add a fifth dimension to the vector. We exploit the vectorial representation of text using a “weighted” version of the pretrained Sentence Transformers\footnote{Sentence Transformers  is a modification of the pretrained BERT network that use siamese and triplets network structures to derive semantically meaningful sentence embeddings that can be compared using cosine-similarity. The main advantage of the model is that it is optimized to calculate cosine similarity, while maintaining the accuracy from BERT. Further information on the architecture is available in the work of \cite{reimers2019sentence}.} embeddings to measure the cosine similarity between the two original names. In particular, considering that pretrained vectors are generic, to refine the vectorial representations for our specific case they have been weighted (token by token) with the result, for each token, of its inverse document frequency (idf)\footnote{The inverse document frequency is the second component of the tf-idf formula. For the purpose of this work, adding the term frequency (tf) component would not be useful, as, in our specific case of company names, any term is unlikely to appear more than once in each name.}, calculated as follows:
\begin{center}
\begin{equation}
 idf= {log\frac{N}{n_{i}}}       
\end{equation}
\end{center}
\medskip
where N the total number of names and $n_{i}$ the number of names in which token \textit{i} appears. Therefore, common tokens have a lower weight compared to non-common ones. The idf score is rescaled in a range between 0 (excluded) and 1 and each token vector is multiplied by its token corresponding idf score. The resulting numerical representation of each company name is then a weighted mean of its token vectors. 
Afterwards, we take the 5th dimensional vector and we compute the scalar product with the vector of weights.\footnote{For \textit{type 1} names the minimum score is -1 and the maximum 5 for the first run of the algorithm.} 

For \textit{type 2} names instead, such as "ADVANCED TECHONOLOGIES INC." or "ENGINEERING PHARMACEUTICAL COMPANY LTD." (11'061 names before preprocessing), we use a bidimensional vector, containing a zero/one if the two companies share the same domain and the measure of their cosine similarity, and we multiply it for the vector of weights.\footnote{For \textit{type 2} names the minimum score is -1 and the maximum 2 for the first run of the algorithm.}

The result for both \textit{type 1} and \textit{type 2} is the matching score for each couple of names in our data. 
In that way we obtain a N*N matrix where each name has been compared with all the remaining (N-1) names and each element of the matrix is the score of the couple.

\subsection{Filtering phase}  \label{subsec4}
\subsubsection{Community detection}
\bigskip{}

\begin{figure}
  \caption{Network visualization}\label{fig:net}
\label{fig:comm}
  \centerline{\includegraphics[width=0.9\textwidth]{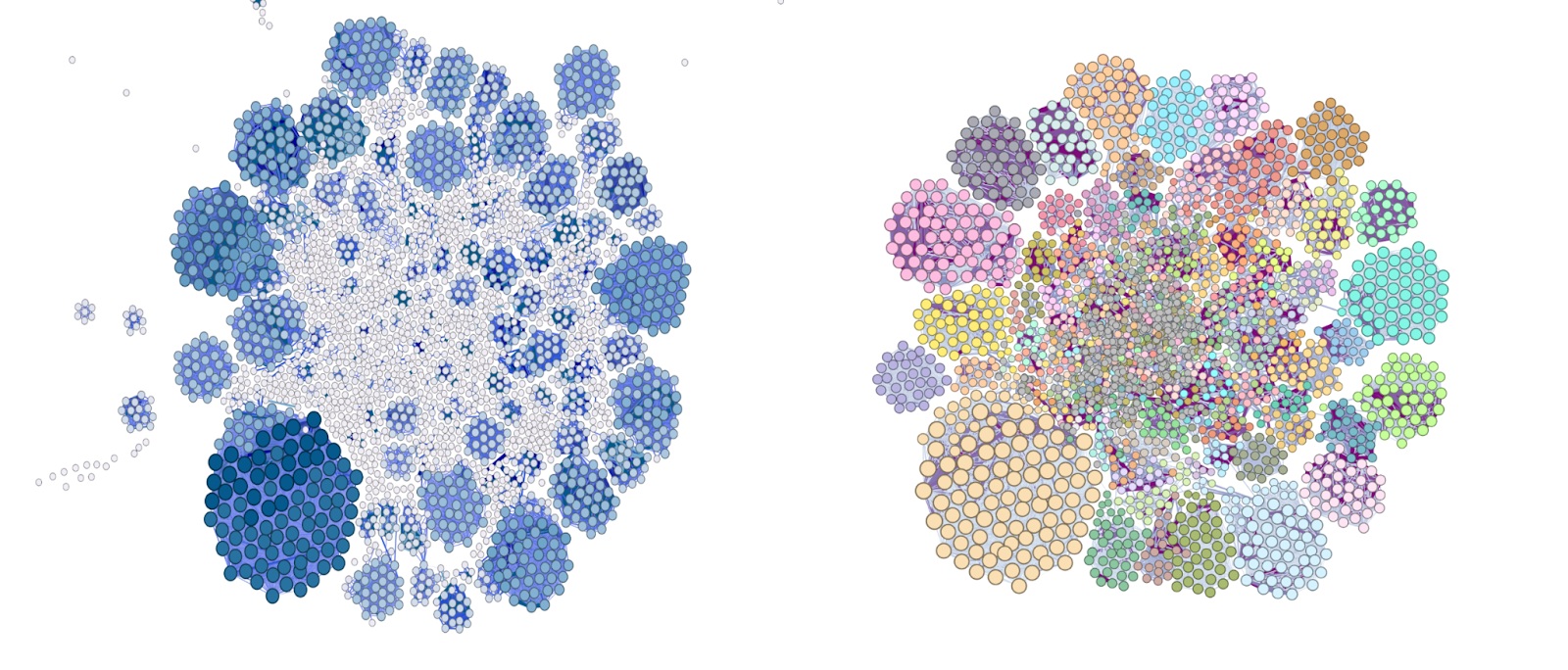}}
\scriptsize
  \justifying{\emph{Notes:} This figure represents the names' network (left) and the names' network after community detection (right); the communities are outlined by different colors.}
\end{figure}

In the filtering stage we set a strategy to identify and reject false positives identified in the matching stage. Despite we also use the filtering stage as a way to select which couples will be considered as referring to the same entity, our filtering stage is very different from the ones proposed, for instance, in \cite{raffo2009play} and in \cite{pezzoni2014kill}. In particular, to decide whether to consider a group of observations as the same entity, we do not follow previous research, that uses entity-related additional filters such as being located in the same country and producing innovation in the same technological field, or patent based filters, such as cross-citations analysis. The \textit{ratio} for this decision is that those filters where conceived to be used on inventors' names, who are unlikely to be the same person when living in two different countries or working in unrelated fields. Therefore, in the case of company names, we believe such filters might be risky to use; for instance, multinational corporations, those presenting the greatest variety of names, have headquarters in different countries and they work in wide technological spaces.
Considering this, to reject false positives we leverage network theory. In particular, we create a network where each name is a node and the edge is present only if the two names' similarity score is over a certain threshold.\footnote{\label{n2}At this point, the threshold is empirically set at 3.9. However, the optimal threshold is also selected in the optimization phase. Refer to the previous note.} The weight of the edge is the score of the previous stage and it is augmented if the two companies have transacted patents in at least a common location.\footnote{The information of the location is available in the \textit{assignee} table in the Patent Assignment Database. In particular, we harmonize the following variables and turn them to single locations: \textit{ee\_city}, \textit{ee\_state}, and \textit{ee\_country} which respectively report city, state, postcode and country related to a certain assignee in a specific transaction. The edges' weights is therefore augmented by 1 point if the two companies share at least one location in which they transacted patents. The empirical weight is also among those optimized later on.}
After creating the network, we use the Louvain algorithm to do community detection, which is a method to extract communities from large networks introduced by \cite{blondel2008fast}.\footnote{\label{n3}By adjusting the \textit{resolution} parameter of the Louvain algorithm, one can control the granularity of the community detection process. A higher resolution parameter may be suitable for detecting fine-grained communities, while a lower resolution parameter may be more appropriate for identifying larger, more cohesive groups \citep{blondel2008fast}. In particular, we empirically set the resolution of the Louvain algorithm at 1 for the current run of Terrorizer and we optimize it later on. We select this value as it is the default option in python \textit{networkx} package\\ (\url{https://networkx.org/documentation/stable/reference/algorithms/generated/networkx.algorithms.community.louvain.louvain_communities.html}), considering that if resolution is less than 1, the algorithm favors larger communities; greater than 1 instead favors smaller communities.
} Figure \ref{fig:net} represents a snapshot of the network before (left) and after community detection with Louvain method; the communities are outlined by different colors (right).

In addition, in order to improve the precision of our results, we perform a further step, removing the edges which stand among communities. We do that by exploiting the idea of \textit{bridgeness centrality} proposed by \cite{jensen2016detecting}.\footnote{\label{n4}For the current run, we delete edges of nodes which have a bridgness centrality greater than 1. The selection of the value is empirical.} The idea of bridgeness centrality is a modification of the concept of betweenness centrality, which has a potential weakness in giving  equal scores to local centers (i.e. nodes of high degree central to a single region) and to global bridges, which connect different communities.
The work of \cite{jensen2016detecting} wants to overcome this limitation, proposing a new measure of centrality which allows us to distinguish between \textit{local} bridges (edges which stand intra-community, which in our case might be, for example,  a shorter version of the names) and \textit{global} bridges (edges which connect in fact different communities), and we delete the latter.\footnote{For a more detailed explanation of the algorithm characteristics, please refer to the paper of \cite{jensen2016detecting}; for the code implementation in python, we used the github repository linked to the paper (\url{https://github.com/mmorini/gSSL})} In our data, the case of global bridges might be reflected in joint ventures names. For example, the joint venture "DEERE-HITACHI" might be linked to both companies Deere and Hitachi, creating misleading bridges among those two different communities. To operationalize this idea, we work in a recursive way: after finding the communities on the total graph, we consider each community as an independent subgraph in which we look for edges with positive bridgeness centrality and delete them. After, on each subgraph, we reapply the Louvain algorithm and find the new communities which are a subset of the starting communities. The same procedure applies to both \textit{type 1} and \textit{type 2} names. 

\subsubsection{Community naming}
\bigskip{}

The second part of the filtering step is to chose a name for each of the identified communities. Consistently with the proposed methodology, we chose the name for each community following a word embedding and names' similarity approach. In particular, our naming strategies has 4 steps:
\begin{enumerate}
    \item Create an encoding (using the Sentence Transformers architecture as explained above \citep{reimers2019sentence}) for each of the names in each community;
    \item Calculate the cosine similarity among all the possible couples;
    \item Calculate the average cosine similarity per group element;
    \item Select, for each community, the name which has the greatest average cosine similarity;
\end{enumerate}
The \textit{ratio} of this naming strategy is to assign to each community the name which is, on average, closer to all the others.
Furthermore, this is not the only naming strategy possible. Another possible strategy could be to assign to each community the name which has more importance in terms of patent volume in our data. In particular, we can calculate the number of patents assigned to each original assignee name and then select for each community the name with the highest number of patents assigned among the companies' names which make up that community.

\subsection{Validation}\label{subsec:val}
\bigskip{}
In order to validate the results obtained applying the \textit{Terrorizer} on USPTO company names, we run the algorithm on four different datasets for which we have the groundtruth. The first two are also used to test the disambiguation algorithm used in PatentsView \citep{monathdisambiguating}\footnote{The two datasets (NBER and PatentsView Assignee) are available at \url{https://data.patentsview.org.s3.amazonaws.com/documents/PatentsView_Disambiguation.pdf}}:
\begin{enumerate}
    \item \textbf{NBER}: The National Bureau of Economic Research provides disambiguated assignee data. These data are created semiautomatically with manual correction and labeling of assignee co-reference decisions produced by string similarity. The NBER assignee dataset has 238’398 number of records, linked to 7’236 distinct entities.
    \item \textbf{PatentsView (PW) assignee}: The PatentsView team created a hand-labeled set of disambiguated assignee records. The data were created by sampling records of each assignee type (universities, federal government entities, private companies, states, and local government agencies) and those records are used as queries for annotators to find all other records referring to the same assignee. Further, this dataset has a larger coverage of name varieties of the entities than the NBER dataset, which is important  to evaluate the more difficult-to-disambiguate cases. It contains 371'599 records referring to 111 distinct entities.
    \item \textbf{Patent Assignment Database random (PAD-R)}: From our dataset of 325'917 companies, we randomly sample  3’657 companies' names. The data are manually labeled in order to add the groundtruth.
    \item \textbf{PAD multinational corporations (PAD-MNC)}; This dataset has been created manually looking for all the names variants for the top 150 companies in terms of patent volume at the USPTO since 2005. Therefore, the resulting dataset contains 8'459 names variants linked to the top 150 assignees.

\end{enumerate}

These gold standard datasets are especially useful as they allow us to assess the performance of Terrorizer and also to compare it on the same data cleaned with a different algorithm (this is the case of NBER and PW assignee datasets, for which we have the results of \cite{monathdisambiguating}). 

For the evaluation, we refer to classic metrics in machine learning: \textit{precision}, \textit{recall} and \textit{F1}. These metrics, explained next, emphasize different aspects of correctness of the clusters that the algorithm creates. Mathematically, they can be defined as follows:

\[
\text{Precision} = \frac{\text{True Positives}}{\text{True Positives + False Positives}}
\]

\[
\text{Recall} = \frac{\text{True Positives}}{\text{True Positives + False Negatives}}
\]

\[
\text{F1} = 2 \times \frac{\text{Precision} \times \text{Recall}}{\text{Precision} + \text{Recall}}
\]

\bigskip{}
where $True Positives$ represent the number of correct positive instances that are correctly identified as positive by the model and $True Negatives$ the number of correct negative instances that are correctly identified as negative by the model. Further, $False Positives$ represent the number of negative instances that are incorrectly identified as positive by the model and $False Negatives$ the number of positive instances that are incorrectly identified as negative by the model.

Table \ref{tab:evaluation} reports the results of the metrics on each dataset. It is worth mentioning that the results from Terrorizer have been obtained with a set of predefined empirical weights and thresholds (see Notes \ref{n1}, \ref{n2}, \ref{n3}, \ref{n4}).

Considering harmonizing company names a multi-label classification problem, the best way to evaluate its performance seems to look at its \textit{micro-average} precision and recall. Indeed, micro-average precision and recall provide a global view of the algorithm's performance across all labels. 
Looking at the results, we have two main findings: the first is that, compared to the algorithm of PW, the F1 is superior, respectively by 0.212 (NBER) and 0.167 (PatentsView Assignee). Looking also at the results obtained on the two original manually samples, we observe that the performance is similar, pointing out the fact that Terrorizer generalizes well on different datasets. 

\begin{landscape}
    \begin{table}[htbp]
\caption{Evaluation Results Comparing the Precision, Recall and F1 Scores for the  PatentsView 2021 Disambiguation Methodology and Terrorizer}
\bigskip{}
\label{tab:evaluation}
\begin{center}
\begin{tabular}{|l|c|c|c|c|}
\hline
\textbf{Metric} & \multicolumn{4}{c|}{\cellcolor[HTML]{EFEFEF}\textbf{Evaluation Datasets}} \\ \cline{2-5} 
 & \textbf{NBER Assignee} & \textbf{PatentsView Assignee} & \textbf{PAD-R} &  \textbf{PAD-MNC} \\ \hline
Precision & 0.355 & 0.518 & \cellcolor[HTML]{EFEFEF}  & \cellcolor[HTML]{EFEFEF} \\ \hline
Recall & 0.508 & 0.555 & \cellcolor[HTML]{EFEFEF}  & \cellcolor[HTML]{EFEFEF} \\ \hline
F1 & 0.419 & 0.536 & \cellcolor[HTML]{EFEFEF} & \cellcolor[HTML]{EFEFEF} \\ \hline
\multicolumn{5}{|l|}{\textbf{Disambiguation methodology for PatentsView data updated through 06/14/2021}} \\ \hline
Precision & 0.559 & 0.643 & 0.617 & 0.989  \\ \hline
Recall & 0.725 & 0.775 & 0.812 & 0.446 \\ \hline
F1 & 0.631 & 0.703 & 0.701 & 0.614 \\ \hline
\multicolumn{5}{|l|}{\textbf{Terrorizer (12/2023 Version)}} \\ \hline
\end{tabular}
  \end{center}

\scriptsize
{Notes: The original NBER assignee  dataset has 238’398 number of records, linked to 7’236 distinct entities (grountruth); however, we manage to recover only names information related to 216’858 linked to 6’176 distinct entities). The original  PatentsView assignee dataset accounts for 371’599 records  linked to 111 unique entities. In this case, we manage to recover only information about 186’238 record linked to 106 unique entities. The PAD assignee dataset is a hand labeled dataset created randomly selecting 3’657 observations from our sample of patent assignees who file at least one utility patent at USPTO from 2005 onwards. Further, all the metrics refer to the \textit{micro-average}.} 

  \end{table}
\end{landscape}

\subsection{Hyperparameter tuning: a Bayesian approach}\label{subsec:opt}
\label{sec:opt}
\bigskip{}

To maximize the performance of Terrorizer, we perform the tuning of its hyperparameters. In machine learning systems, hyperparameter tuning is a critical step to ensure that the system performs optimally, efficiently, and relevantly to the problem at hand. In the case of Terrorizer, we decide to perform the hyperparameter optimization on the procedure to harmonize \textit{type 1} names, as they constitute more than 90\% of the total number of names. Our goal is to maximize the performance of Terrorizer in terms of F1 score. 
In particular, the above presented methodology has a number of hyperparameters:

\begin{enumerate}
    \item The weights of the matching phase; \footnote{In particular, the weights refer to following conditions: the two names have a token in common, the first token is in common, the domain extracted from the first url is the same, the two names have a token in common in the text of the first url. Further, we tune also the weight to be assigned to the cosine similarity.}
    \item The threshold of the matching phase;
    \item The threshold of the bridgeness centrality;
    \item The resolution of Louvain's algorithm;
    \item The weight to be added to the edges when two companies have at least one location in common;
\end{enumerate}

For a total of 9 hyperparameters. In the research we use the Optuna framework to perform hyperparameter optimization \citep{akiba2019optuna}. This technique provides a more accurate parameter search compared to grid search by efficiently pruning suboptimal parameter combinations and continuously refining the search space. Optuna is known for its easy way of dynamically constructing complex hyperparameters' optimization search spaces by its so-called Define-by-run API. Optuna employs, among others, the Tree-structured Parzen Estimator (TPE), a Bayesian optimization method, widely used in recent parameter tuning frameworks \citep{tree}. In general, a hyperparameter tuning process aims at minimizing/maximizing an objective function $f(\theta)$, where $\theta$ represents the hyperparameters. Mathematically, the optimization process is represented as:
\begin{equation}
    \theta^{*}=\arg \max_{\theta} f(\theta)
\end{equation}

Each trial in Optuna proposes a new set of hyperparameters $\theta$, in order to find the optimal hyperparameters $\theta^{*}$ that minimize/maximize the objective function across trials that take into account the results of the previous ones. Through this systematic and guided approach, Optuna helped enhance our algorithm's performance by identifying a refined set of hyperparameters.\\

We run the optimization process across all our benchmark datasets, in order to see the metrics when the hyperparameters are optimized. In particular, we want to maximize the F1 score. Table \ref{tab:hyper} in Appendix \ref{sec:appB} reports the results on each dataset. However, we decide to use the hyperparameters' values obtained from the dataset NBER described in the subsection \ref{subsec:val}. We chose this dataset because it is the largest in terms of size, which can potentially contribute to more generalizable hyperparameters, provided the dataset is also diverse and representative of the problem space.

\section{Names disambiguation results}

\label{sec:res}
\begin{figure}
\caption{No. of unique assignees year by year (original vs Terrorizer)}
\bigskip{}
\label{fig:gres}

  \centerline{\includegraphics[width=0.9\textwidth]{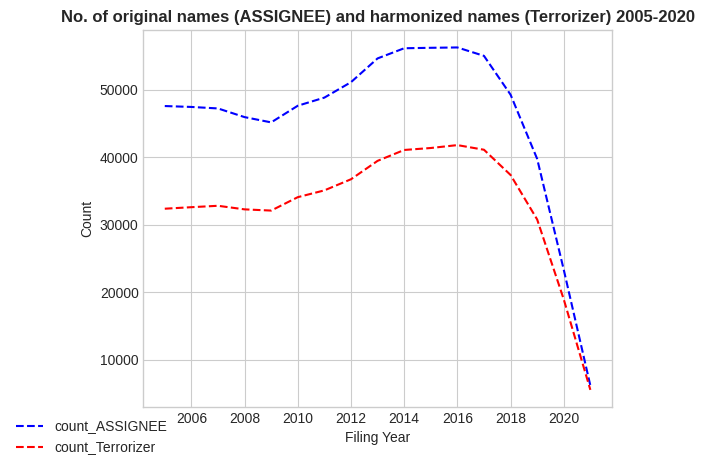}}
\scriptsize
  \justifying{\emph{Notes:} The figure shows the number of unique assignees per year in the time range 2005-2020 using the original USPTO names and the harmonized names.}
\label{fig:result}
\end{figure}

Figure \ref{fig:gres} reports the number of unique original USPTO assignees ($count\_ASSIGNEE$) and the number of unique assignees after using the Terrorizer algorithm ($count\_Terrorizer$). The overall reduction of names is around 42\%.
In particular, Terrorizer entity resolution method works in a variety of cases including the following:
\begin{enumerate}
\item \textit{geographical indications}: Terrorizer manages to harmonize the different geographical indications which are included in the names of the same entity. For instance, "NOKIA DENMARK A/D", "NOKIA US HOLDING" and "NOKIA CANADA INC." are all assigned to the "NOKIA" community. At the same time, "BASF (CHINA) COMPANY LIMITED", "BASF AUSTRALIA LTD.", "BASF ANTWERPEN N.V." , "BASF ESPANOLA S.L." are all assigned to the entity "BASF".
\item \textit{business units and subsidiaries}: Terrorizer is able to link to the same business entity the different departments of the same company and its subsidiaries (legal entity harmonization). For example, "IBM JAPAN BUSINESS LOGISTICS CO., LTD." and "IBM GLOBAL SERVICES PTE. LTD." are both linked to the entity "IBM"; "SONY MOBILE COMMUNICATIONS JAPAN, INC." and "SONY ELECTRONICS INC., A DELAWARE CORPORATION" and "SONY CHEMICALS CORPORATION OF AMERICA" are all linked to the entity "SONY".
\item \textit{legal forms}: Terrorizer can harmonize different legal indications which refer to the same entity. For example, "MICHELIN RECHERCHE ET TEHCNIQUE, S.A." and "SOCIETE DE TECHNOGIE MICHELIN" are both linked to the "MICHELIN" entity; "PFIZER LIMITED", "PFIZER, INC.", "PFIZER ITALIA S.R.L." and "PFIZER ENTERPRISES SARL" are all linked to the company "PFIZER".
\item \textit{acronyms}: Terrorizer might also recognize acronyms and link them to the correct company. For instance, in USPTO data we find "INTERNATIONAL BUSINESS MACHINES COR" and "INTERNATIONAL BUSINESS MACHINES COMPANY" which are both linked to the company "IBM".
\end{enumerate}

\bigskip{}

In addition to reducing the number of unique names referring to the same company, Terrorizer's results also have an impact on the problem of the underestimation of patent portfolios, which is characteristic of companies with many name variants. Table \ref{tab:uniqass} shows for the top 10 companies in our data, both in terms of patent volume and the number of associated name variants, the differences in terms of unique names and patent portfolio size before and after the use of Terrorizer. Among the top 150 companies in terms of patent volume, we select the 10 names with the largest number of variants associated with them. The number of variants is reported in column 4 - "Name var." (Name Variants). In particular, column 4  reports the number of unique names' variants we were able to manually identify in our data as referring to that same entity through manual work.\footnote{In particular, for each company we manually search our assignee list for all possible name variants of this company. For this purpose, we select the most characteristic token(s) for each name, e.g. "NOKIA", to identify the variants of the company Nokia. In this way, we obtain a list of names containing the characterizing token(s), which we manually check to exclude false positives (such as the company Nokian in the case of Nokia). We also manually check for each company whether there are versions of the characterizing token that contain spelling errors in order to achieve maximum completeness. However, we cannot be certain that we actually capture all name variants present in our data.} The company "HITACHI" is ranked first, with 424 variants, followed by the company "SAMSUNG" with 325 variants. The first column "Company" reports the official name of the company. The second column "Most used assignee name" contains the name variant of a particular company that has the most patents associated with it in our data compared to the other variants. For instance, for the company "HITACHI", the most frequent variant is "HITACHI, LTD". The third column "Portfolio size" shows the number of patents belonging to the company name variant in the second column (that is, the variant name "HITACHI, LTD." is linked to 11'175 patents). The fifth column, "Names after Terr." (Names after Terrorizer), reports instead the number of unique names after using Terrorizer. Finally, the last column "Port. size after Terr." (Portfolio size after Terrorizer), reports the number of patents belonging to the cluster, generated by Terrorizer, where the name in the second column is contained.

In our data, the company "HITACHI" exhibits many of the problems associated with the task of entity linking presented in Section \ref{section2}. In particular, some variants have spelling errors, such as "HITACHI HIGH-TECHMOLOGIES CORPORATION" and the company has several divisions, including "HITACHI HIGH-TECH CORPORATION", "HITACHI INFORMATION \& CONTROL SOLUTIONS, LTD." and "HITACHI INFORMATION \& COMMUNICATION ENGINEERING. LTD.". In addition, Hitachi has many subsidiaries in our data, such as "HITACHI ULSI SYSTEMS CO. LTD" and "HITACHI UNISIA AUTOMOTIVE, LTD.". As it is a Japanese company, we also find related names of Hitachi in Japanese, such as "KYUSHU HITACHI, LTD.". The geographical indications associated with the company name include cases such as "HITACHI AUTOMOTIVE SYSTEMS EUROPE GMBH", "HITACHI DATA STORAGE KOREA, INC." and "HITACHI CHEMICAL CO. AMERICA, LTD.". With Terrorizer, we manage to reduce the number of unique names by 84\% and increase the number of patents associated with the Hitachi cluster by 50\%. In other cases, the number of patents increases even more, as in the case of the Sumitomo company, where it increases by 224\%, or in the case of Tyco, where it increases by 185\%. In other cases, such as Sony, the number of patents increases only slightly despite a 77\% decrease in the number of unique names. This is possible because many of Sony's name variants end up in a different cluster than where the name "SONY CORPORATION" is.

\begin{landscape}
\begin{table}[htbp]
\begin{center}
\caption{Patent portfolio changes for major companies}
\label{tab:uniqass}
\medskip
\setlength{\tabcolsep}{5pt} 
\renewcommand{\arraystretch}{1.2}
\begin{tabular}{|l|l|l|l|l|l|}
\hline
\rowcolor[HTML]{EFEFEF} 
\textbf{Company} & \textbf{Most used assignee name} & \textbf{Portfolio} & \textbf{Name} & \textbf{Names} & \textbf{Port. size} \\
\rowcolor[HTML]{EFEFEF} 
& & \textbf{size} & \textbf{var.} & \textbf{Terr.} & \textbf{after Terr.} \\ \hline

HITACHI & HITACHI, LTD. & 11'175 & 424 & 67 & 16'736 \\ \hline
SAMSUNG & SAMSUNG ELECTRONICS CO., LTD. & 69'085 & 395 & 70 & 75'153 \\ \hline
MITSUBISHI & MITSUBISHI ELECTRIC CORPORATION & 11'697 & 266 & 17 & 22'464 \\ \hline
AT\&T INTELLECTUAL PROPERTY & AT\&T INTELLECTUAL PROPERTY I, L.P. & 8ì307 & 233 & 40 & 12'648 \\ \hline
SONY & SONY CORPORATION & 30'245 & 216 & 49 & 30'297 \\ \hline
DOW & DOW GLOBAL TECHNOLOGIES LLC & 2'954 & 212 & 2 & 5'751 \\ \hline
TOSHIBA & KABUSHIKI KAISHA TOSHIBA & 26'134 & 204 & 13 & 31'539 \\ \hline
TYCO & TYCO HEALTHCARE GROUP LP & 3'241 & 187 & 7 & 9'220 \\ \hline
GENERAL ELECTRIC COMPANY & GENERAL ELECTRIC COMPANY & 22'545 & 182 & 55 & 23'106 \\ \hline
SUMITOMO & SUMITOMO ELECTRIC INDUSTRIES, LTD. & 4'455 & 176 & 3 & 14'461 \\ \hline
\end{tabular}
\end{center}
\end{table}
\scriptsize
{Notes: The table presents a comprehensive overview of patent assignments for major companies, detailing the original number of name variations, the reduced number after applying the Terrorizer clustering, and the total patents assigned before and after this process in our data. The first column "Company" reports the name of the company; the second column "Most used assignee name" reports the variant, of a certain company, linked to the greatest patent volume. The third column "Portfolio size" shows the number of patents belonging to the company name variant in the second column.  The fourth column "Name var." (Name Variants) reports instead the number of unique names' variants we were able to manually identify in our data as referring to that same entity. The fifth column, "Names after Terr." (Names after Terrorizer), reports instead the number of unique names after using Terrorizer. Finally, the last column "Port. size after Terr." (Portfolio size after Terrorizer), reports the number of patents belonging to the cluster, generated by Terrorizer, where the name in the second column is included.}

\end{landscape}

\section{Conclusion, limitations and further research}
\label{sec:co}

Terrorizer is a company name harmonization algorithm that has shown promising results in disambiguating variations of company names and other entities in USPTO patent data. Compared to its predecessors, it proposes a novel methodology that goes beyond hard-coded solutions and string matching techniques and utilizes a mixture of the latest NLP deep learning techniques, together with a rule-based approach and network theory principles.

Compared to other entity liking algorithms that use a "three-phase method", such as in \cite{raffo2009play} and in \cite{pezzoni2014kill}, Terrorizer has several original elements. First, a knowledge augmentation phase is added, which uses the information accessible via queries in a web browser to augment knowledge about company names. This step is crucial to the methodology as it allows even \textit{adversarial} cases such as the companies NOKIAN and NOKIA, which are different entities despite semantic similarity, and to harmonize cases such as "BASF EAST ASIA REGIONAL HEADQUARTERS LTD." and "BASF ARGICULTURAL SOLUTIONS SEED US LLC", which are indeed characterized by major semantic differences. This type of result cannot be achieved using conventional string matching techniques.
Second, Terrorizer uses a network approach to form communities and identify company names belonging to the same entity through community generation. The main advantage of this type of filtering is that it allows refinement and aggregation of results using techniques from network theory, so that additional matches can be excluded or created that would not be possible by comparing textual information alone.

Terrorizer's results are stable across different datasets, even if they contain entities that are not companies, such as universities, hospitals and government institutions. If one compares the performance of Terrorizer directly with the algorithm proposed by \cite{monathdisambiguating}, Terrorizer performs better in terms of F1 score.

\bigskip{}
However, this work is not free from limitations. First of all, Terrorizer's effectiveness is to some extent limited to those companies for which at least some kind of information is available on the Internet, i.e. it is likely that Terrorizer's performance is better for larger companies, such as multinational corporations, than for very small companies. In addition, Terrorizer also leverages the reliability of the rankings suggested by the search engines. It should be noted that these rankings change over time and this cannot be controlled by people using the information displayed. This could lead to instability in Terrorizer's results over time.
Another weakness of Terrorizer is that it is not possible to reconstruct the changes in ownership of companies over time. In this sense, Terrorizer groups together companies that might have separated at a different point in time, or vice versa. For this type of work, access to private data sources is crucial  to achieve good results, as demonstrated in the recent work by \cite{arora2021matching}. However, this is beyond the current scope of Terrorizer.
Furthermore, the benchmark datasets used to evaluate its performance and optimize the hyperparameters of Terrorizer could be improved. In particular, given their size, which is less than 20\% of the size of the data we apply the algorithm to, they may not reflect the complexity of the actual data well. This leads to two different problems: the first is a problem in evaluating the performance of Terrorizer, while the second is that the optimized hyperparameters may not lead to better performance of the algorithm when using such a small amount of data for the tuning process.

\bigskip{}
Future improvements to Terrorizer could also extend the matching phase to other types of attributes, such as company logos or the Legal Entity Identifier (LEI). In particular, the LEI code can be used to identify named entities even when they undergo different types of legal transactions, such as mergers, acquisitions, consolidations, purchases and management takeovers. In this case, the company name may change after the legal transaction while still referring to the same entity \citep{basile2024disambiguation}.
Another interesting avenue to explore would be to use the results of Terrorizer to train an NLP model. From this perspective, Terrorizer can be seen as a step in a continuous learning process in which the results are used as input for training more sophisticated similarity models.
\clearpage

\bibliographystyle{elsarticle-num-names}
\bibliography{arxiv_terrorizer}

\begin{thebibliography}{43}
\expandafter\ifx\csname natexlab\endcsname\relax\def\natexlab#1{#1}\fi
\providecommand{\url}[1]{\texttt{#1}}
\providecommand{\href}[2]{#2}
\providecommand{\path}[1]{#1}
\providecommand{\DOIprefix}{doi:}
\providecommand{\ArXivprefix}{arXiv:}
\providecommand{\URLprefix}{URL: }
\providecommand{\Pubmedprefix}{pmid:}
\providecommand{\doi}[1]{\href{http://dx.doi.org/#1}{\path{#1}}}
\providecommand{\Pubmed}[1]{\href{pmid:#1}{\path{#1}}}
\providecommand{\bibinfo}[2]{#2}
\ifx\xfnm\relax \def\xfnm[#1]{\unskip,\space#1}\fi
\bibitem[{Monath et~al.(2021)Monath, Madhavan, DiPietro, McCallum, and Jones}]{monathdisambiguating}
\bibinfo{author}{N.~Monath}, \bibinfo{author}{S.~Madhavan}, \bibinfo{author}{C.~DiPietro}, \bibinfo{author}{A.~McCallum}, \bibinfo{author}{C.~Jones},
\newblock \bibinfo{title}{Disambiguating patent inventors, assignees, and their locations in patentsview},
\newblock \bibinfo{journal}{Tech. Rep}  (\bibinfo{year}{2021}).
\bibitem[{Griliches(1998)}]{griliches1998patent}
\bibinfo{author}{Z.~Griliches},
\newblock \bibinfo{title}{Patent statistics as economic indicators: a survey},
\newblock in: \bibinfo{booktitle}{R\&D and productivity: the econometric evidence}, \bibinfo{publisher}{University of Chicago Press}, \bibinfo{year}{1998}, pp. \bibinfo{pages}{287--343}.
\bibitem[{Abraham and Moitra(2001)}]{abraham2001innovation}
\bibinfo{author}{B.~P. Abraham}, \bibinfo{author}{S.~D. Moitra},
\newblock \bibinfo{title}{Innovation assessment through patent analysis},
\newblock \bibinfo{journal}{Technovation} \bibinfo{volume}{21} (\bibinfo{year}{2001}) \bibinfo{pages}{245--252}.
\bibitem[{Hall and Harhoff(2012)}]{hall2012recent}
\bibinfo{author}{B.~H. Hall}, \bibinfo{author}{D.~Harhoff},
\newblock \bibinfo{title}{Recent research on the economics of patents},
\newblock \bibinfo{journal}{Annu. Rev. Econ.} \bibinfo{volume}{4} (\bibinfo{year}{2012}) \bibinfo{pages}{541--565}.
\bibitem[{Sampat(2018)}]{sampat2018survey}
\bibinfo{author}{B.~N. Sampat},
\newblock \bibinfo{title}{A survey of empirical evidence on patents and innovation},
\newblock \bibinfo{journal}{NBER WORKING PAPER SERIES}  (\bibinfo{year}{2018}).
\bibitem[{Pavitt(1985)}]{pavitt1985patent}
\bibinfo{author}{K.~Pavitt},
\newblock \bibinfo{title}{Patent statistics as indicators of innovative activities: possibilities and problems},
\newblock \bibinfo{journal}{Scientometrics} \bibinfo{volume}{7} (\bibinfo{year}{1985}) \bibinfo{pages}{77--99}.
\bibitem[{Von~Wartburg et~al.(2005)Von~Wartburg, Teichert, and Rost}]{von2005inventive}
\bibinfo{author}{I.~Von~Wartburg}, \bibinfo{author}{T.~Teichert}, \bibinfo{author}{K.~Rost},
\newblock \bibinfo{title}{Inventive progress measured by multi-stage patent citation analysis},
\newblock \bibinfo{journal}{research Policy} \bibinfo{volume}{34} (\bibinfo{year}{2005}) \bibinfo{pages}{1591--1607}.
\bibitem[{Van~Looy et~al.(2006)Van~Looy, Du~Plessis, and Magerman}]{van2006data}
\bibinfo{author}{B.~Van~Looy}, \bibinfo{author}{M.~Du~Plessis}, \bibinfo{author}{T.~Magerman},
\newblock \bibinfo{title}{Data production methods for hamonized patent statistics: Patentee sector allocation},
\newblock \bibinfo{journal}{Available at SSRN 944464}  (\bibinfo{year}{2006}).
\bibitem[{Squicciarini et~al.(2013)Squicciarini, Dernis, and Criscuolo}]{squicciarini2013measuring}
\bibinfo{author}{M.~Squicciarini}, \bibinfo{author}{H.~Dernis}, \bibinfo{author}{C.~Criscuolo},
\newblock \bibinfo{title}{Measuring patent quality: Indicators of technological and economic value},
\newblock \bibinfo{journal}{OECD Science, Technology and Industry Working Papers}  (\bibinfo{year}{2013}).
\bibitem[{Magerman et~al.(2006)Magerman, Van~Looy, and Song}]{magerman2006data}
\bibinfo{author}{T.~Magerman}, \bibinfo{author}{B.~Van~Looy}, \bibinfo{author}{X.~Song},
\newblock \bibinfo{title}{Data production methods for harmonized patent statistics: Patentee name harmonization},
\newblock \bibinfo{journal}{Available at SSRN 944470}  (\bibinfo{year}{2006}).
\bibitem[{Raffo and Lhuillery(2009)}]{raffo2009play}
\bibinfo{author}{J.~Raffo}, \bibinfo{author}{S.~Lhuillery},
\newblock \bibinfo{title}{How to play the “names game”: Patent retrieval comparing different heuristics},
\newblock \bibinfo{journal}{Research policy} \bibinfo{volume}{38} (\bibinfo{year}{2009}) \bibinfo{pages}{1617--1627}.
\bibitem[{Carayol et~al.(2009)Carayol, Cassi et~al.}]{carayol2009s}
\bibinfo{author}{N.~Carayol}, \bibinfo{author}{L.~Cassi}, et~al.,
\newblock \bibinfo{title}{Who’s who in patents. a bayesian approach},
\newblock \bibinfo{journal}{Cahiers du GREThA} \bibinfo{volume}{7} (\bibinfo{year}{2009}) \bibinfo{pages}{07--2009}.
\bibitem[{Thoma et~al.(2010)Thoma, Torrisi, Gambardella, Guellec, Hall, and Harhoff}]{thoma2010harmonizing}
\bibinfo{author}{G.~Thoma}, \bibinfo{author}{S.~Torrisi}, \bibinfo{author}{A.~Gambardella}, \bibinfo{author}{D.~Guellec}, \bibinfo{author}{B.~H. Hall}, \bibinfo{author}{D.~Harhoff}, \bibinfo{title}{Harmonizing and combining large datasets-An application to firm-level patent and accounting data}, \bibinfo{type}{Technical Report}, National Bureau of Economic Research, \bibinfo{year}{2010}.
\bibitem[{Peeters et~al.(2010)Peeters, Song, Callaert, Grouwels, and Van~Looy}]{peeters2010harmonizing}
\bibinfo{author}{B.~Peeters}, \bibinfo{author}{X.~Song}, \bibinfo{author}{J.~Callaert}, \bibinfo{author}{J.~Grouwels}, \bibinfo{author}{B.~Van~Looy},
\newblock \bibinfo{title}{Harmonizing harmonized patentee names: an exploratory assessment of top patentees},
\newblock \bibinfo{journal}{Eurostat Working Paper}  (\bibinfo{year}{2010}).
\bibitem[{Pezzoni et~al.(2014)Pezzoni, Lissoni, and Tarasconi}]{pezzoni2014kill}
\bibinfo{author}{M.~Pezzoni}, \bibinfo{author}{F.~Lissoni}, \bibinfo{author}{G.~Tarasconi},
\newblock \bibinfo{title}{How to kill inventors: testing the massacrator{\copyright} algorithm for inventor disambiguation},
\newblock \bibinfo{journal}{Scientometrics} \bibinfo{volume}{101} (\bibinfo{year}{2014}) \bibinfo{pages}{477--504}.
\bibitem[{Morrison et~al.(2017)Morrison, Riccaboni, and Pammolli}]{morrison2017disambiguation}
\bibinfo{author}{G.~Morrison}, \bibinfo{author}{M.~Riccaboni}, \bibinfo{author}{F.~Pammolli},
\newblock \bibinfo{title}{Disambiguation of patent inventors and assignees using high-resolution geolocation data},
\newblock \bibinfo{journal}{Scientific data} \bibinfo{volume}{4} (\bibinfo{year}{2017}) \bibinfo{pages}{1--21}.
\bibitem[{Gay et~al.(2008)Gay, Latham, and Le~Bas}]{gay2008collective}
\bibinfo{author}{C.~Gay}, \bibinfo{author}{W.~Latham}, \bibinfo{author}{C.~Le~Bas},
\newblock \bibinfo{title}{Collective knowledge, prolific inventors and the value of inventions: An empirical study of french, german and british patents in the us, 1975--1999},
\newblock \bibinfo{journal}{Economics of Innovation and New Technology} \bibinfo{volume}{17} (\bibinfo{year}{2008}) \bibinfo{pages}{5--22}.
\bibitem[{Trajtenberg et~al.(2006)Trajtenberg, Shiff, and Melamed}]{trajtenberg2006names}
\bibinfo{author}{M.~Trajtenberg}, \bibinfo{author}{G.~Shiff}, \bibinfo{author}{R.~Melamed}, \bibinfo{title}{The" names game": Harnessing inventors' patent data for economic research}, \bibinfo{year}{2006}.
\bibitem[{Lai et~al.(2009)Lai, D’Amour, and Fleming}]{lai2009careers}
\bibinfo{author}{R.~Lai}, \bibinfo{author}{A.~D’Amour}, \bibinfo{author}{L.~Fleming},
\newblock \bibinfo{title}{The careers and co-authorship networks of us patent-holders, since 1975},
\newblock \bibinfo{journal}{Unpublished Working Paper, Harvard University}  (\bibinfo{year}{2009}).
\bibitem[{Migu{\'e}lez and G{\'o}mez-Migu{\'e}lez(2011)}]{miguelez2011singling}
\bibinfo{author}{E.~Migu{\'e}lez}, \bibinfo{author}{I.~G{\'o}mez-Migu{\'e}lez},
\newblock \bibinfo{title}{Singling out individual inventors from patent data},
\newblock \bibinfo{journal}{Available at SSRN 1856875}  (\bibinfo{year}{2011}).
\bibitem[{Li et~al.(2014)Li, Lai, D’Amour, Doolin, Sun, Torvik, Amy, and Fleming}]{li2014disambiguation}
\bibinfo{author}{G.-C. Li}, \bibinfo{author}{R.~Lai}, \bibinfo{author}{A.~D’Amour}, \bibinfo{author}{D.~M. Doolin}, \bibinfo{author}{Y.~Sun}, \bibinfo{author}{V.~I. Torvik}, \bibinfo{author}{Z.~Y. Amy}, \bibinfo{author}{L.~Fleming},
\newblock \bibinfo{title}{Disambiguation and co-authorship networks of the us patent inventor database (1975--2010)},
\newblock \bibinfo{journal}{Research Policy} \bibinfo{volume}{43} (\bibinfo{year}{2014}) \bibinfo{pages}{941--955}.
\bibitem[{Ventura et~al.(2015)Ventura, Nugent, and Fuchs}]{ventura2015seeing}
\bibinfo{author}{S.~L. Ventura}, \bibinfo{author}{R.~Nugent}, \bibinfo{author}{E.~R. Fuchs},
\newblock \bibinfo{title}{Seeing the non-stars:(some) sources of bias in past disambiguation approaches and a new public tool leveraging labeled records},
\newblock \bibinfo{journal}{Research Policy} \bibinfo{volume}{44} (\bibinfo{year}{2015}) \bibinfo{pages}{1672--1701}.
\bibitem[{Kim et~al.(2016)Kim, Khabsa, and Giles}]{kim2016inventor}
\bibinfo{author}{K.~Kim}, \bibinfo{author}{M.~Khabsa}, \bibinfo{author}{C.~L. Giles},
\newblock \bibinfo{title}{Inventor name disambiguation for a patent database using a random forest and dbscan},
\newblock in: \bibinfo{booktitle}{Proceedings of the 16th ACM/IEEE-CS on Joint Conference on Digital Libraries}, \bibinfo{year}{2016}, pp. \bibinfo{pages}{269--270}.
\bibitem[{Han et~al.(2019)Han, Yu, Wang, Zhai, Ran, and Han}]{han2019disambiguating}
\bibinfo{author}{H.~Han}, \bibinfo{author}{Y.~Yu}, \bibinfo{author}{L.~Wang}, \bibinfo{author}{X.~Zhai}, \bibinfo{author}{Y.~Ran}, \bibinfo{author}{J.~Han},
\newblock \bibinfo{title}{Disambiguating uspto inventor names with semantic fingerprinting and dbscan clustering},
\newblock \bibinfo{journal}{The Electronic Library} \bibinfo{volume}{37} (\bibinfo{year}{2019}) \bibinfo{pages}{225--239}.
\bibitem[{Petrie and Julius(2023)}]{petrie2023novel}
\bibinfo{author}{S.~M. Petrie}, \bibinfo{author}{T.~D. Julius},
\newblock \bibinfo{title}{A novel text representation which enables image classifiers to also simultaneously classify text, applied to name disambiguation},
\newblock \bibinfo{journal}{Scientometrics}  (\bibinfo{year}{2023}) \bibinfo{pages}{1--25}.
\bibitem[{Neuh{\"a}usler et~al.(2017)Neuh{\"a}usler, Frietsch, Mund, and Eckl}]{neuhausler2017identifying}
\bibinfo{author}{P.~Neuh{\"a}usler}, \bibinfo{author}{R.~Frietsch}, \bibinfo{author}{C.~Mund}, \bibinfo{author}{V.~Eckl},
\newblock \bibinfo{title}{Identifying the technology profiles of r\&d performing firms—a matching of r\&d and patent data},
\newblock \bibinfo{journal}{International Journal of Innovation and Technology Management} \bibinfo{volume}{14} (\bibinfo{year}{2017}) \bibinfo{pages}{1740003}.
\bibitem[{Lee et~al.(2023)Lee, Park, and Lee}]{lee2023simple}
\bibinfo{author}{J.~Lee}, \bibinfo{author}{S.~Park}, \bibinfo{author}{J.~Lee},
\newblock \bibinfo{title}{Simple and effective way to disambiguate and standardize patent applicants using an attention mechanism with data augmentation},
\newblock \bibinfo{journal}{IEEE Access}  (\bibinfo{year}{2023}).
\bibitem[{Arora et~al.(2021)Arora, Belenzon, and Sheer}]{arora2021matching}
\bibinfo{author}{A.~Arora}, \bibinfo{author}{S.~Belenzon}, \bibinfo{author}{L.~Sheer},
\newblock \bibinfo{title}{Matching patents to compustat firms, 1980--2015: Dynamic reassignment, name changes, and ownership structures},
\newblock \bibinfo{journal}{Research Policy} \bibinfo{volume}{50} (\bibinfo{year}{2021}) \bibinfo{pages}{104217}.
\bibitem[{Onishi et~al.(2012)Onishi, Nishimura, Tsukada, Yamauchi, Shimbo, Kani, and Nakamura}]{onishi2012standardization}
\bibinfo{author}{K.~Onishi}, \bibinfo{author}{Y.~Nishimura}, \bibinfo{author}{N.~Tsukada}, \bibinfo{author}{I.~Yamauchi}, \bibinfo{author}{T.~Shimbo}, \bibinfo{author}{M.~Kani}, \bibinfo{author}{K.~Nakamura},
\newblock \bibinfo{title}{Standardization and accuracy of japanese patent applicant names},
\newblock \bibinfo{journal}{Available at SSRN 2147190}  (\bibinfo{year}{2012}).
\bibitem[{Li et~al.(2018)Li, Hu, Cui, and Hu}]{li2018deeppatent}
\bibinfo{author}{S.~Li}, \bibinfo{author}{J.~Hu}, \bibinfo{author}{Y.~Cui}, \bibinfo{author}{J.~Hu},
\newblock \bibinfo{title}{Deeppatent: patent classification with convolutional neural networks and word embedding},
\newblock \bibinfo{journal}{Scientometrics} \bibinfo{volume}{117} (\bibinfo{year}{2018}) \bibinfo{pages}{721--744}.
\bibitem[{Sarica et~al.(2020)Sarica, Luo, and Wood}]{sarica2020technet}
\bibinfo{author}{S.~Sarica}, \bibinfo{author}{J.~Luo}, \bibinfo{author}{K.~L. Wood},
\newblock \bibinfo{title}{Technet: Technology semantic network based on patent data},
\newblock \bibinfo{journal}{Expert Systems with Applications} \bibinfo{volume}{142} (\bibinfo{year}{2020}) \bibinfo{pages}{112995}.
\bibitem[{Trappey et~al.(2019)Trappey, Chen, Trappey, and Ma}]{trappey2019machine}
\bibinfo{author}{A.~J. Trappey}, \bibinfo{author}{P.~P. Chen}, \bibinfo{author}{C.~V. Trappey}, \bibinfo{author}{L.~Ma},
\newblock \bibinfo{title}{A machine learning approach for solar power technology review and patent evolution analysis},
\newblock \bibinfo{journal}{Applied Sciences} \bibinfo{volume}{9} (\bibinfo{year}{2019}) \bibinfo{pages}{1478}.
\bibitem[{Arts et~al.(2021)Arts, Hou, and Gomez}]{arts2021natural}
\bibinfo{author}{S.~Arts}, \bibinfo{author}{J.~Hou}, \bibinfo{author}{J.~C. Gomez},
\newblock \bibinfo{title}{Natural language processing to identify the creation and impact of new technologies in patent text: Code, data, and new measures},
\newblock \bibinfo{journal}{Research Policy} \bibinfo{volume}{50} (\bibinfo{year}{2021}) \bibinfo{pages}{104144}.
\bibitem[{Graham et~al.(2018)Graham, Marco, and Myers}]{graham2018patent}
\bibinfo{author}{S.~J. Graham}, \bibinfo{author}{A.~C. Marco}, \bibinfo{author}{A.~F. Myers},
\newblock \bibinfo{title}{Patent transactions in the marketplace: Lessons from the uspto patent assignment dataset},
\newblock \bibinfo{journal}{Journal of Economics \& Management Strategy} \bibinfo{volume}{27} (\bibinfo{year}{2018}) \bibinfo{pages}{343--371}.
\bibitem[{Sanh et~al.(2019)Sanh, Debut, Chaumond, and Wolf}]{sanh2019distilbert}
\bibinfo{author}{V.~Sanh}, \bibinfo{author}{L.~Debut}, \bibinfo{author}{J.~Chaumond}, \bibinfo{author}{T.~Wolf},
\newblock \bibinfo{title}{Distilbert, a distilled version of bert: smaller, faster, cheaper and lighter},
\newblock \bibinfo{journal}{arXiv preprint arXiv:1910.01108}  (\bibinfo{year}{2019}).
\bibitem[{Autor et~al.(2020)Autor, Dorn, Hanson, Pisano, and Shu}]{autor2020foreign}
\bibinfo{author}{D.~Autor}, \bibinfo{author}{D.~Dorn}, \bibinfo{author}{G.~H. Hanson}, \bibinfo{author}{G.~Pisano}, \bibinfo{author}{P.~Shu},
\newblock \bibinfo{title}{Foreign competition and domestic innovation: Evidence from us patents},
\newblock \bibinfo{journal}{American Economic Review: Insights} \bibinfo{volume}{2} (\bibinfo{year}{2020}) \bibinfo{pages}{357--374}.
\bibitem[{Reimers and Gurevych(2019)}]{reimers2019sentence}
\bibinfo{author}{N.~Reimers}, \bibinfo{author}{I.~Gurevych},
\newblock \bibinfo{title}{Sentence-bert: Sentence embeddings using siamese bert-networks},
\newblock \bibinfo{journal}{arXiv preprint arXiv:1908.10084}  (\bibinfo{year}{2019}).
\bibitem[{Blondel et~al.(2008)Blondel, Guillaume, Lambiotte, and Lefebvre}]{blondel2008fast}
\bibinfo{author}{V.~D. Blondel}, \bibinfo{author}{J.-L. Guillaume}, \bibinfo{author}{R.~Lambiotte}, \bibinfo{author}{E.~Lefebvre},
\newblock \bibinfo{title}{Fast unfolding of communities in large networks},
\newblock \bibinfo{journal}{Journal of statistical mechanics: theory and experiment} \bibinfo{volume}{2008} (\bibinfo{year}{2008}) \bibinfo{pages}{P10008}.
\bibitem[{Jensen et~al.(2016)Jensen, Morini, Karsai, Venturini, Vespignani, Jacomy, Cointet, Merckl{\'e}, and Fleury}]{jensen2016detecting}
\bibinfo{author}{P.~Jensen}, \bibinfo{author}{M.~Morini}, \bibinfo{author}{M.~Karsai}, \bibinfo{author}{T.~Venturini}, \bibinfo{author}{A.~Vespignani}, \bibinfo{author}{M.~Jacomy}, \bibinfo{author}{J.-P. Cointet}, \bibinfo{author}{P.~Merckl{\'e}}, \bibinfo{author}{E.~Fleury},
\newblock \bibinfo{title}{Detecting global bridges in networks},
\newblock \bibinfo{journal}{Journal of Complex Networks} \bibinfo{volume}{4} (\bibinfo{year}{2016}) \bibinfo{pages}{319--329}.
\bibitem[{Akiba et~al.(2019)Akiba, Sano, Yanase, Ohta, and Koyama}]{akiba2019optuna}
\bibinfo{author}{T.~Akiba}, \bibinfo{author}{S.~Sano}, \bibinfo{author}{T.~Yanase}, \bibinfo{author}{T.~Ohta}, \bibinfo{author}{M.~Koyama},
\newblock \bibinfo{title}{Optuna: A next-generation hyperparameter optimization framework},
\newblock in: \bibinfo{booktitle}{Proceedings of the 25th ACM SIGKDD international conference on knowledge discovery \& data mining}, \bibinfo{year}{2019}, pp. \bibinfo{pages}{2623--2631}.
\bibitem[{Watanabe(2023)}]{tree}
\bibinfo{author}{S.~Watanabe}, \bibinfo{title}{Tree-structured parzen estimator: Understanding its algorithm components and their roles for better empirical performance}, \bibinfo{year}{2023}. \href{http://arxiv.org/abs/2304.11127}{{\tt arXiv:2304.11127}}.
\bibitem[{Basile et~al.(2024)Basile, Crupi, Grasso, Mercanti, Regoli, Scarsi, Yang, and Cosentini}]{basile2024disambiguation}
\bibinfo{author}{A.~Basile}, \bibinfo{author}{R.~Crupi}, \bibinfo{author}{M.~Grasso}, \bibinfo{author}{A.~Mercanti}, \bibinfo{author}{D.~Regoli}, \bibinfo{author}{S.~Scarsi}, \bibinfo{author}{S.~Yang}, \bibinfo{author}{A.~C. Cosentini},
\newblock \bibinfo{title}{Disambiguation of company names via deep recurrent networks},
\newblock \bibinfo{journal}{Expert Systems with Applications} \bibinfo{volume}{238} (\bibinfo{year}{2024}) \bibinfo{pages}{122035}.
\bibitem[{Coffano and Tarasconi(2014)}]{coffano2014crios}
\bibinfo{author}{M.~Coffano}, \bibinfo{author}{G.~Tarasconi},
\newblock \bibinfo{title}{Crios-patstat database: sources, contents and access rules},
\newblock \bibinfo{journal}{Center for Research on Innovation, Organization and Strategy, CRIOS working paper}  (\bibinfo{year}{2014}).

\end{thebibliography}

\clearpage
\section*{Appendices}

\renewcommand{\thesubsection}{\Alph{subsection}}

\subsection{Appendix I}\label{sec:SectionA.3}

\renewcommand{\thetable}{A\arabic{table}}
\setcounter{figure}{0}
\renewcommand{\thefigure}{A\arabic{figure}}
\section*{Finetuning DistilRoBERTa to identify patent assignees' kind}

In this research we want to perform entity resolution on patent assignees that are company name. Consistently with that purpose, we first need to effectively distinguish organization names from individual names among patent assignees, as previously done in \cite{coffano2014crios}. The mentioned approach consists in individuating a proper name if the considered string exactly matches the following structure "Name, Surname". However, this idea brings a series of limitations, considering not all the individual names follow this semantic structure, as they can be names that do not present the comma and also individual names that include more than one name or surname. To improve the current approach, we create a binary classifier using DistilRoBERTa (6-layer, 768- hidden, 12-heads, 82B of parameters.)
DistilRoBERTa is a variant of the RoBERTa (A Robustly Optimized BERT Pretraining Approach) model that has been distilled for efficiency while retaining strong language understanding capabilities \citep{sanh2019distilbert}\footnote{The implemented model is freely accessible at \url{https://huggingface.co/distilroberta-base}}.
To fine tune the model, we use a manually labeled datasets of 19'398 patent assignees names from PatentsView\footnote{We start from the \textit{g\_assignee\_disambiguated} table and manually add cases of individual names which do not matches the "Name, Surname" pattern.}, of which 10'000 are labeled as organization names and 9'398 are labeled instead as individual names. In particular, we allocate 60\% of the data to the training set and 40\% to the validation set. Then, we further split the validation data in validation and test set, allocating 50\% of the observations to each of them; therefore, test and validation have 25\% of the original observations each. We apply stratified splitting to maintain the same distribution of labels for each set. We train the classifier for 3 epochs, using a batch size of 32. The selected loss function is the Cross Entropy Loss. For the 0 class (organization) we achieve a precision of 0.98 and a recall of 0.97. For the 1 class (individual) we achieve a precision of 0.97 and a recall of 0.98.
\clearpage
\subsection{Appendix II}\label{sec:appB}
\setcounter{table}{0}
\section*{Hyperparameter optimization}

\begin{table}[htbp]
\caption{Hyperparameter optimization using TPE on gold standards}
\bigskip{}
\label{tab:hyper}
\begin{center}
\begin{tabular}{|l|c|c|c|c|}
\hline
\textbf{Hyperparameter} & \multicolumn{4}{c|}{\cellcolor[HTML]{EFEFEF}\textbf{ Datasets}} \\ \cline{2-5} 
 & \textbf{NBER Assignee} & \textbf{PatentsView Assignee} & \textbf{PAD-R} &  \textbf{PAD-MNC} \\ \hline
First token & 0.49 & 0.40 & 0.72 & 0.45 \\ \hline
Token in common & 0.17 & 0.56 & 0.99 & 0.17 \\ \hline
Domain & 0.81 & 0.93 & 0.95 & 0.28 \\ \hline
Cosine similarity & 0.70 & 0.61 & 0.97 & 0.65 \\ \hline
Text in first url & 0.52 & 0.79 & 0.54 & 1.80 \\ \hline
Threshold matching & 1.80 & 2.34 & 0.94 & 0.50 \\ \hline
Resolution (Louvain) & 0.50 & 0.33 & 0.45 & 1.80 \\ \hline
Bridgeness centrality & 1.80 & 0.50 & 1.06 & 0.71 \\ \hline
\textbf{Optimized F1} & \textbf{0.71} & \textbf{0.75} & \textbf{0.93} & \textbf{0.93} \\ \hline
\end{tabular}
\end{center}

\scriptsize
{Notes: There are a total of 8 hyperparameters: the first five refer to the \textit{matching stage}: first token, token in common, domain, cosine similarity, text in the first url. Then we have the threshold of the matching stage (Threshold matching). For the network phase we have: resolution (Louvain) and bridgeness centrality. The search spaces for each of these hyperparameters are, respectively: 0.1,1 for the five matching score components, 0.5,5 for the matching threshold, 0.001, 0.1 for the resolution and -2 to 2 for the bridgeness centrality. The empirical run of Terrorizer used the following empirically defined weights: 1 for each component of the matching phase, 3.9 for the matching score, 1 for the resolution and for the bridgeness centrality. Further, the F1 score refers to the \textit{micro-average}.} 
\end{table}




\end{document}